%%%%%%%%%%%%%%%%%%%%%%%%%%%%%%%%%%%%%%%%%%%%%%%%%%%%%%%%%%%%%%%%%%%%%%%%%%%%%%%%
%                                                                              %
%       Pouliot Type Duality via a-Maximization                                %
%                                                                              %
%    T. Kawano, Y. Ookouchi, Y. Tachikawa, and F. Yagi  (University of Tokyo)  %
%                                                                              %
%%%%%%%%%%%%%%%%%%%%%%%%%%%%%%%%%%%%%%%%%%%%%%%%%%%%%%%%%%%%%%%%%%%%%%%%%%%%%%%%

%\def\mydraft{My Draft!}
%\def\mydraft{b }

\input harvmac
\input tables 
\input epsf.tex

\noblackbox

%% Macro
%style
\def\ie{{\it i.e.},$\ $}
%general
\def\hpt#1{{\tt {#1}}}

%Macro for figure
\newcount\figno
\figno=0
\def\fig#1#2#3{
\par\begingroup\parindent=0pt\leftskip=1cm\rightskip=1cm\parindent=0pt
\baselineskip=11pt
\global\advance\figno by 1
\midinsert
\epsfxsize=#3
\centerline{\epsfbox{#2}}
\vskip 12pt
{\it Figure \the\figno:} #1\par
\endinsert\endgroup\par
}
\def\figlabel#1{\xdef#1{\the\figno}}

%%%%% NEW DEF OOKOUCHI%%%%%%%%
\def\le{\leq}
\def\ge{\geq}
\def\ele{{\rm ele}}
\def\mag{{\rm mag}}
%%%%%%%%%%%%%%%%%%%%%%%%%%%%%%%

% ==========================================================================
% Young tableaux macros
% ==========================================================================

%1
\def\fund{  \> {\vcenter  {\vbox  
              {\hrule height.6pt
               \hbox {\vrule width.6pt  height5pt  
                      \kern5pt 
                      \vrule width.6pt  height5pt }
               \hrule height.6pt}
                         }
                   }
           \>\> }

%1bar
\def\antifund{  \> \overline{ {\vcenter  {\vbox  
              {\hrule height.6pt
               \hbox {\vrule width.6pt  height5pt  
                      \kern5pt 
                      \vrule width.6pt  height5pt }
               \hrule height.6pt}
                         }
                   } }
           \>\> }

%2
\def\sym{  \> {\vcenter  {\vbox  
              {\hrule height.6pt
               \hbox {\vrule width.6pt  height5pt  
                      \kern5pt 
                      \vrule width.6pt  height5pt 
                      \kern5pt
                      \vrule width.6pt height5pt}
               \hrule height.6pt}
                         }
              }
           \>\> }

%2bar
\def\symbar{  \> \overline{ {\vcenter  {\vbox  
              {\hrule height.6pt
               \hbox {\vrule width.6pt  height5pt  
                      \kern5pt 
                      \vrule width.6pt  height5pt 
                      \kern5pt
                      \vrule width.6pt height5pt}
               \hrule height.6pt}
                         }
              }
           } \>\> }

%11
\def\anti{ \> {\vcenter  {\vbox  
              {\hrule height.6pt
               \hbox {\vrule width.6pt  height5pt  
                      \kern5pt 
                      \vrule width.6pt  height5pt }
               \hrule height.6pt
               \hbox {\vrule width.6pt  height5pt  
                      \kern5pt 
                      \vrule width.6pt  height5pt }
               \hrule height.6pt}
                         }
              }
           \>\> }

%111
\def\antithree{ \> 
              {\vcenter  {\vbox  
              {\hrule height.6pt
               \hbox {\vrule width.6pt  height5pt  
                      \kern5pt 
                      \vrule width.6pt  height5pt }
               \hrule height.6pt
               \hbox {\vrule width.6pt  height5pt  
                      \kern5pt 
                      \vrule width.6pt  height5pt }
               \hrule height.6pt
               \hbox {\vrule width.6pt  height5pt  
                      \kern5pt 
                      \vrule width.6pt  height5pt }
               \hrule height.6pt}
                         }
              }
           \>\> }

%1111
\def\antifour{ \> 
              {\vcenter  {\vbox  
              {\hrule height.6pt
               \hbox {\vrule width.6pt  height5pt  
                      \kern5pt 
                      \vrule width.6pt  height5pt }
               \hrule height.6pt
               \hbox {\vrule width.6pt  height5pt  
                      \kern5pt 
                      \vrule width.6pt  height5pt }
               \hrule height.6pt
               \hbox {\vrule width.6pt  height5pt  
                      \kern5pt 
                      \vrule width.6pt  height5pt }
               \hrule height.6pt
               \hbox {\vrule width.6pt  height5pt  
                      \kern5pt 
                      \vrule width.6pt  height5pt }
               \hrule height.6pt}
                         }
              }
           \>\> }

%11111
\def\antifive{ \> 
              {\vcenter  {\vbox  
              {\hrule height.6pt
               \hbox {\vrule width.6pt  height5pt  
                      \kern5pt 
                      \vrule width.6pt  height5pt }
               \hrule height.6pt
               \hbox {\vrule width.6pt  height5pt  
                      \kern5pt 
                      \vrule width.6pt  height5pt }
               \hrule height.6pt
               \hbox {\vrule width.6pt  height5pt  
                      \kern5pt 
                      \vrule width.6pt  height5pt }
               \hrule height.6pt
               \hbox {\vrule width.6pt  height5pt  
                      \kern5pt 
                      \vrule width.6pt  height5pt }
               \hrule height.6pt
               \hbox {\vrule width.6pt  height5pt  
                      \kern5pt 
                      \vrule width.6pt  height5pt }
               \hrule height.6pt}
                         }
              }
           \>\> }

%22
\def\twotwo{
              {\vcenter  {\vbox
              {\hrule height.5pt
               \hbox {\vrule width.5pt  height4pt
                      \kern4pt
                      \vrule width.5pt  height4pt
                      \kern4pt
                      \vrule width.5pt height4pt}
               \hrule height.5pt
               \hbox {\vrule width.5pt  height4pt
                      \kern4pt
                      \vrule width.5pt  height4pt
                      \kern4pt
                      \vrule width.5pt height4pt}
               \hrule height.5pt}
                         }
              }
           \>\> }

%4
\def\four{  {\vcenter  {\vbox
              {\hrule height.5pt
               \hbox {\vrule width.5pt  height4pt
                      \kern4pt
                      \vrule width.5pt  height4pt
                      \kern4pt
                      \vrule width.5pt  height4pt
                      \kern4pt
                      \vrule width.5pt  height4pt
                      \kern4pt
                      \vrule width.5pt height4pt}
               \hrule height.5pt}
                         }
              }
           }

%%%%%%%%%%%%%%%%%%%%%%%%%%%%%%%

% References

\lref\PSX{
  P.~Pouliot and M.~J.~Strassler,
  ``Duality and Dynamical Supersymmetry Breaking in $Spin(10)$ with a Spinor,''
  Phys.\ Lett.\ B {\bf 375}, 175 (1996), 
  \hpt{hep-th/9602031}.
  %%CITATION = HEP-TH 9602031;%%
}

\lref\kawano{
  T.~Kawano,
  ``Duality of ${\cal N}=1$ Supersymmetric $SO(10)$ Gauge Theory with Matter 
    in the Spinorial Representation,''
  Prog.\ Theor.\ Phys.\  {\bf 95}, 963 (1996), 
  \hpt{hep-th/9602035}.
  %%CITATION = HEP-TH 9602035;%%
}

\lref\SpinX{
M.~Berkooz, P.~L.~Cho, P.~Kraus and M.~J.~Strassler,
  ``Dual Descriptions of $SO(10)$ SUSY Gauge Theories with Arbitrary Numbers  of
  Spinors and Vectors,''
  Phys.\ Rev.\ D {\bf 56}, 7166 (1997)
  \hpt{hep-th/9705003}.
  %%CITATION = HEP-TH 9705003;%%
}

\lref\PSVIII{
  P.~Pouliot and M.~J.~Strassler,
  ``A Chiral $SU(N)$ Gauge Theory and its Non-Chiral $Spin(8)$ Dual,''
  Phys.\ Lett.\ B {\bf 370}, 76 (1996), 
  \hpt{hep-th/9510228}.
  %%CITATION = HEP-TH 9510228;%%
}

\lref\ADS{
  I.~Affleck, M.~Dine and N.~Seiberg,
  ``Exponential Hierarchy from Dynamical Supersymmetry Breaking,''
  Phys.\ Lett.\ B {\bf 140}, 59 (1984).
  %%CITATION = PHLTA,B140,59;%%
}

\lref\murayama{
  H.~Murayama,
  ``Studying Noncalculable Models of Dynamical Supersymmetry Breaking,''
  Phys.\ Lett.\ B {\bf 355}, 187 (1995), 
  \hpt{hep-th/9505082}.
  %%CITATION = HEP-TH 9505082;%%
}

\lref\pouliot{
  P.~Pouliot,
  ``Chiral Duals of Nonchiral SUSY Gauge Theories,''
  Phys.\ Lett.\ B {\bf 359}, 108 (1995)
  \hpt{hep-th/9507018}.
  %%CITATION = HEP-TH 9507018;%%
}

\lref\EM{
  N.~Seiberg,
  ``Electric-Magnetic Duality in Supersymmetric NonAbelian Gauge Theories,''
  Nucl.\ Phys.\ B {\bf 435}, 129 (1995)
  \hpt{hep-th/9411149}.
  %%CITATION = HEP-TH 9411149;%%
}

\lref\Mack{
  G.~Mack,
  ``All Unitary Ray Representations of the Conformal Group $SU(2,2)$ with 
  Positive Energy,''
  Commun.\ Math.\ Phys.\  {\bf 55}, 1 (1977).
  %%CITATION = CMPHA,55,1;%%
}

\lref\athm{
  E.~Barnes, K.~Intriligator, B.~Wecht and J.~Wright,
  ``Evidence for the Strongest Version of the 4d $a$-Theorem, via $a$-Maximization 
  along RG Flows,''
  Nucl.\ Phys.\ B {\bf 702}, 131 (2004)
  \hpt{hep-th/0408156}.
  %%CITATION = HEP-TH 0408156;%%
}

\lref\bound{
  D.~Kutasov, A.~Parnachev and D.~A.~Sahakyan,
  ``Central Charges and $U(1)_R$ Symmetries in ${\cal N} = 1$ Super Yang-Mills,''
  JHEP {\bf 0311}, 013 (2003)
  \hpt{hep-th/0308071}.
  %%CITATION = HEP-TH 0308071;%%
}

\lref\amax{
  K.~Intriligator and B.~Wecht,
  ``The Exact Superconformal $R$-Symmetry Maximizes $a$,''
  Nucl.\ Phys.\ B {\bf 667}, 183 (2003), 
  \hpt{hep-th/0304128}.
  %%CITATION = HEP-TH 0304128;%%
}

\lref\tHooft{
  G.~'t Hooft,
  ``Naturalness, Chiral Symmetry, and Spontaneous Chiral Symmetry Breaking,''
    in {\it Recent Developments in Gauge Theories}, eds. 't~Hooft 
    {\it et.$\,$al.} (Plenum Press, New York, 1980), 135. 
}

\lref\unitarity{
  D.~Kutasov and A.~Schwimmer,
  ``On Duality in Supersymmetric Yang-Mills Theory,''
  Phys.\ Lett.\ B {\bf 354}, 315 (1995)
  \hpt{hep-th/9505004}.
  %%CITATION = HEP-TH 9505004;%%
}

\lref\BGISW{
  E.~Barnes, E.~Gorbatov, K.~Intriligator, M.~Sudano and J.~Wright,
  ``The Exact Superconformal $R$-Symmetry Minimizes $\tau_{RR}$,''
  \hpt{hep-th/0507137}.
  %%CITATION = HEP-TH 0507137;%%
}

\lref\Tachikawa{
  Y.~Tachikawa,
  ``Five-Dimensional Supergravity Dual of $a$-Maximization,''
  \hpt{hep-th/0507057}.
  %%CITATION = HEP-TH 0507057;%%
}

\lref\Cardy{
  J.~L.~Cardy,
  ``Is There a $C$ Theorem in Four Dimensions?,''
  Phys.\ Lett.\ B {\bf 215}, 749 (1988).
  %%CITATION = PHLTA,B215,749;%%
}

\lref\IWADE{
K.~Intriligator and B.~Wecht,
``RG Fixed Points and Flows in SQCD with Adjoints,''
Nucl.\ Phys.\ B {\bf 677}, 223 (2004)
\hpt{hep-th/0309201}.
%%CITATION = HEP-TH 0309201;%%
}

\lref\AnselmiI{
D.~Anselmi, J.~Erlich, D.~Z.~Freedman and A.~A.~Johansen,
``Positivity Constraints on Anomalies in Supersymmetric Gauge
Theories,''
Phys.\ Rev.\ D {\bf 57}, 7570 (1998)
 \hpt{hep-th/9711035}.
%%CITATION = HEP-TH 9711035;%%
}

\lref\AnselmiII{
D.~Anselmi, D.~Z.~Freedman, M.~T.~Grisaru and A.~A.~Johansen,
``Nonperturbative Formulas for Central Functions of Supersymmetric Gauge
Theories,''
Nucl.\ Phys.\ B {\bf 526}, 543 (1998)
\hpt{hep-th/9708042}.
%%CITATION = HEP-TH 9708042;%%
}

\lref\KOTY{
  T.~Kawano, Y.~Ookouchi and F.~Yagi, 
  in preparation.
}

\lref\Adler{
  S.~L.~Adler,
  ``Axial Vector Vertex in Spinor Electrodynamics,''
  Phys.\ Rev.\  {\bf 177}, 2426 (1969)
  %%CITATION = PHRVA,177,2426;%%
}

\lref\BellJackiw{
  J.~S.~Bell and R.~Jackiw,
  ``A PCAC Puzzle: $\pi_0 \to \gamma\gamma$ in the Sigma Model,''
  Nuovo Cim.\ A {\bf 60}, 47 (1969).
  %%CITATION = NUCIA,A60,47;%%
}

\lref\Keni{
  K.~Intriligator, 
  ``IR Free or Interacting? A Proposed Diagnostic,'' 
  UCSD-PTH-05-14, 
  \hpt{hep-th/0509085}.
}

\lref\OkOo{
  T.~Okuda and Y.~Ookouchi, 
  ``Higgsing and Superpotential Deformations of ADE Superconformal Theories,'' 
  CALT-68-2574, UT-05-11, 
  \hpt{hep-th/0508189}.
  %%CITATION = HEP-TH 0508189;%%
}

\lref\IINSY{
 M.~Ibe, Izawa~K.-I., Y.~Nakayama, Y.~Shinbara, T.~Yanagida, 
 ``More on Conformally Sequestered SUSY Breaking,''
 UT-05-13, {\tt hep-ph/0509229}.
}

\lref\CDSW{
  F.~Cachazo, M.~R.~Douglas, N.~Seiberg and E.~Witten,
  ``Chiral Rings and Anomalies in Supersymmetric Gauge Theory,''
  JHEP {\bf 0212}, 071 (2002), 
  \hpt{hep-th/0211170}.
  %%CITATION = HEP-TH 0211170;%%
}

\lref\AddF{
  N.~Seiberg,
  ``Adding Fundamental Matter to `Chiral Rings and Anomalies in Supersymmetric 
  Gauge Theory',''
  JHEP {\bf 0301}, 061 (2003), 
  \hpt{hep-th/0212225}.
  %%CITATION = HEP-TH 0212225;%%
}

%% the end of references

%%%%%%%%%%%%%%%%%%%%%%%%%%%%%%%%%%%%%%%%%%%%%%%%%%%%%%%%%%%%%%%%%
%                      Title Page                               %
%%%%%%%%%%%%%%%%%%%%%%%%%%%%%%%%%%%%%%%%%%%%%%%%%%%%%%%%%%%%%%%%%
\Title{                                \vbox{\hbox{UT-05-12}
					     \hbox{NSF-KITP-05-75} } }
{\vbox{\centerline{
             Pouliot Type Duality via $a$-Maximization
}}}

\vskip .2in

\centerline{
Teruhiko~Kawano$^1$, Yutaka~Ookouchi$^1$, Yuji~Tachikawa$^{1,2}$, and Futoshi~Yagi$^1$
}

\vskip .2in 

%\centerline{
%      Department of Physics, University of Tokyo, Hongo, Tokyo
%}

\centerline{${}^1$ \sl
               Department of Physics, University of Tokyo,
}
\centerline{\sl
                     Hongo, Tokyo 113-0033, Japan
}
\bigskip

\centerline{${}^2$ \sl 
               Kavli Institute for Theoretical Physics,
}
\centerline{\sl
                    University of California,  Santa Barbara, CA 93106, USA
}

%\centerline{\tt
%   kawano,$\,$ookouchi,$\,$yujitach,$\,$yagi@hep-th.phys.s.u-tokyo.ac.jp
%}
%\vskip -0.05in
%\centerline{\tt
%                    ookouchi@hep-th.phys.s.u-tokyo.ac.jp
%}

\vskip 2cm
%%%%%%%%%%%%%%%%%%%%%%%%%%%%%%%%%%%%%%%%%%%%%%%%%%%%%%%%%%%%%%%%%%%%%%%%%%%%%
% Abstract                                                                  %
%%%%%%%%%%%%%%%%%%%%%%%%%%%%%%%%%%%%%%%%%%%%%%%%%%%%%%%%%%%%%%%%%%%%%%%%%%%%%
\noindent

We study four-dimensional ${\cal N}=1$ $Spin$(10) gauge theory with 
a single spinor and $N_Q$ vectors at the superconformal fixed point 
via the electric-magnetic duality and $a$-maximization. 
When gauge invariant chiral primary operators hit the unitarity bounds, 
we find that the theory with no superpotential is identical to 
the one with some superpotential at the infrared fixed point. 
%The dual pair of the former flows into the one of the latter in the deepest 
%infrared. 
The auxiliary field method in the electric theory offers a satisfying 
description of the infrared fixed point, which is consistent with the better 
picture in the magnetic theory. In particular, it gives a clear description of 
the emergence of new massless degrees of freedom in the electric theory.  

\bigskip\bigskip
\Date{September, 2005}

%%%%%%%%%%%%%%%%%%%%%%%%%%%%%%%%%%%%%%%%%%%%%%%%%%%%%%%%%%%%%%%%%%
%                        CONTENT                                 %
%%%%%%%%%%%%%%%%%%%%%%%%%%%%%%%%%%%%%%%%%%%%%%%%%%%%%%%%%%%%%%%%%%

%%%%%%%%%%%%%%%%%%%%%%%%%%%%%%%%%%%%%%%%%%%%%%%%%%%%%%%%%%%%%%%%%%%%%%
\newsec{Introduction}

Four-dimensional ${\cal N}=1$ $Spin$(10) gauge theory with one chiral superfield in the spinor representation and $N_Q$ chiral superfields in the vector 
representation has rich and intriguing dynamics. In particular, it shows 
dynamical supersymmetry breaking \refs{\ADS,\murayama} with no vectors  
and the electric-magnetic duality \refs{\PSX,\kawano} for 
$7\leq{N}_Q\leq21$, the latter of which leads 
via the gauge symmetry breaking at some points in the moduli space 
to the duality \pouliot\ between chiral and vector-like gauge theories, 
as well as the one discussed in \PSVIII. They all are so called 
the Pouliot-type dualities. 

When the electric-magnetic duality is available, the dual pair is often 
found in the non-Abelian Coulomb phase \EM. Since the theory is at 
the non-trivial infrared fixed point, some exact results can be obtained 
by ${\cal N}=1$ superconformal symmetry. In particular, 
the scaling dimension $D({\cal O})$ of a gauge invariant chiral primary 
operator ${\cal O}$ can be determined by the $U(1)_R$ charge $R({\cal O})$ as
$${
D({\cal O})={3\over2}R({\cal O}). 
}$$
The unitarity of representations of conformal symmetry requires 
the scaling dimension $D({\cal O})$ of a scalar field ${\cal O}$ to satisfy 
\Mack
$${
D({\cal O}) \ge 1.
}$$
However, one sometimes encounters a gauge invariant chiral primary spinless 
operator ${\cal O}$ which appears to satisfy the inequality $R({\cal O})<2/3$. 
It has been discussed that such an operator ${\cal O}$ decouples as a free field 
from the remaining interacting system, and an accidental $U(1)$ symmetry 
appears in the infrared to fix the $U(1)_R$ charge of the operator 
${\cal O}$ to $2/3$ \refs{\EM,\unitarity,\bound}. 

One can see in the paper \pouliot\ that one of the examples is 
$Spin$(7) gauge theory with $N_f=7$ spinors $Q^i$ ($i=1,\cdots,N_f$) and 
with no superpotential. 
Its dual or magnetic theory exists for $7\le{N_f}\le14$ 
and is given by $SU(N_f-4)$ gauge theory with $N_f$ 
antifundamentals $\bar{q}_i$ and a single symmetric tensor $s$, along with 
gauge singlets $M^{ij}$, which can be identified with $Q^iQ^j$ in the electric 
theory. The superpotential $W_{\rm mag}$ of the magnetic theory is given by
$${
W_{\rm mag}={\tilde{h}\over\tilde\mu^2}M^{ij}\bar{q}_i\,{s}\,\bar{q}_j
+{1\over\tilde\mu^{N_f-7}}\det{s},
}$$
where $\tilde\mu$ is a dimensionful parameter to give the correct mass dimension
to $M^{ij}$, and the dimensionless parameter $\tilde{h}$ shows up 
because we assume that the field $M^{ij}$ has the canonical kinetic term. 

As discussed in \pouliot, since the $U(1)_R$ charge of the spinors $Q^i$ 
is given by $1-(5/N_f)$, the gauge invariant operator $M^{ij}$ appears to 
violate the unitarity bound for $N_f=7$ and therefore propagates as a free field at the infrared fixed point. From the viewpoint of the magnetic theory, 
it implies that the parameter $\tilde{h}$ in the superpotential $W_{\rm mag}$ 
goes to zero in the infrared. Then, the F-term condition 
$${
{\tilde{h}\over\tilde\mu^2}\,\bar{q}_i\,s\,\bar{q}_j
={\partial W_{\rm mag} \over \partial M^{ij}}=0
}$$
doesn't impose any constraints on the gauge invariant operators 
$N_{ij}=\bar{q}_i\,s\,\bar{q}_j$, and the new massless degrees of freedom $N_{ij}$ show up in the low-energy spectrum. One can easily see that the resulting magnetic theory at the fixed point 
has a different electric dual from the original electric theory with no 
superpotential. 

In fact, its electric dual is the same as the original electric theory except 
that it has the non-zero superpotential 
$${
W_{\ele}={1\over\mu}N_{ij}Q^iQ^j,
}$$
along with free singlets $M^{ij}$. Thus, one can conclude that these two electric theories are identical at the 
infrared fixed point. It also means that the original dual pair, 
consisting of the $Spin(7)$ gauge theory with no superpotential and the 
magnetic theory with the superpotential $W_{\rm mag}$, flows into another dual 
pair, consisting of the $Spin(7)$ theory with the superpotential $W_{\ele}$ 
and the magnetic dual with vanishing $\tilde{h}$ in the superpotential in the 
deepest infrared. 

From the point of view on the electric side, the same dynamics can be captured 
by the auxiliary field method\foot{This method has been employed in \athm\ 
to give a more elaborate argument about the prescription in \bound\ to 
give the trial $a$-function when gauge invariant operators hit the unitarity 
bounds. Here and also below, we extend the idea for the models under 
consideration in this paper.}.
In the original $Spin(7)$ gauge theory, 
turning on the superpotential 
$${
W={1\over\mu}N_{ij}\left(Q^iQ^j-{h}\,M^{ij}\right),
}$$
where the auxiliary fields $M^{ij}$ and the Lagrange multipliers 
$N_{ij}$ are introduced with the parameter $h$, does not change the original 
theory at all, as far as $h$ is non-zero. The equations of motion give 
the constraints 
$${
Q^iQ^j=h\,M^{ij}, \qquad h\,N_{ij}=0.
}$$
In the case $N_f=7$, since the $U(1)_R$ charge of the operator $M^{ij}$ hits 
the unitarity bound, the interaction of the field $M^{ij}$ vanishes. 
Therefore, the coupling $h$ goes to zero in the infrared to be consistent 
with the magnetic picture. 
When $h=0$, it apparently becomes a different theory from the original 
one and gives the above-mentioned $Spin(7)$ theory with the superpotential 
$W_{\ele}$. In addition, the F-term condition means that the 
directions $Q^iQ^j$ are redundant, but the new degrees of freedom 
$N_{ij}$ are gained. 
%This is in full agreement with the magnetic picture. 
Thus, the auxiliary field method 
gives a satisfying description of gauge invariant operators hitting the 
unitarity bounds on the electric side. 

So far, we have seen that the data of the $U(1)_R$ charges is very powerful 
to uncover the rich infrared dynamics at the superconformal fixed points. 
However, when a superconformal theory has global $U(1)$ 
symmetries  other than $U(1)_R$ symmetry, 
there {\it a priori} exists difficulty in finding which linear combination of 
$U(1)$ symmetries belongs to the superconformal algebra, 
as in our $Spin(10)$ gauge theory. 
This is the place that $a$-maximization \amax\ 
comes to the rescue. The application of the $a$-maximization method to our 
$Spin(10)$ theory is one of the main points of this paper, where we have 
one flavor $U(1)$ symmetry other than the $U(1)_R$ symmetry. 

Let us suppose there are several non-anomalous flavor 
$U(1)$ symmetries other than $U(1)_\lambda$ symmetry. 
The latter transforms gaugino 
$\lambda_\alpha$ as $\lambda_\alpha\to{e}^{i\theta}\lambda_\alpha$ 
and, if necessary to make it non-anomalous, 
the other fields in an appropriate way\foot{Here, just for 
simplicity, we assume that the gauge group is simple.}, while the former 
leaves the gaugino intact.  
%(in terms of the field strength superfield $W_\alpha(\theta)$, 
%$W_\alpha(\theta)\to{e}^{ia}W_\alpha({e}^{-ia}\theta)$). 
The superconformal $U(1)_R$ symmetry, if it isn't an accidental symmetry 
in the infrared, should be given by a linear combination of these $U(1)$ 
symmetries. Therefore, the $U(1)_R$ charge $R_I$ of an operator $\Phi_I$ 
in the infrared may be given by the flavor $U(1)$ charges $F^i{}_I$ 
and the $U(1)_\lambda$ charge $\Lambda_I$ as 
$R_I(s)=\Lambda_I+\sum_{i}s^i{}F^i{}_I$ with fixed real numbers $s^i$, 
where the index $i$ labels the $U(1)$ symmetries other than the $U(1)_\lambda$ 
symmetry. Since the $U(1)_R$ current and the energy-momentum tensor belong to 
the same superconformal multiplet, the anomaly coefficient in the three-point 
function with one of the flavor $U(1)$ currents inserted at one vertex and 
the $U(1)_R$ current at each of the two remaining vertices is related to 
the one with the same flavor $U(1)$ current at one vertex and 
the energy-momentum tensor at each of the remaining two vertices in the 
corresponding triangle Feynman diagrams. 
Therefore, as Intriligator and Wecht discussed in the seminal paper \amax, 
the above parameters $s^i$ giving the superconformal $U(1)_R$ symmetry 
are required to be the solution to 
\eqn\amaximization{
{\partial~ \over \partial s^i}\,a(s)=0, 
\qquad 
{\partial^2~ \over \partial s^i\,\partial s^j}\,a(s)<0, 
}
for all $i,\,j$, where the function $a(s)$ is given 
\refs{\AnselmiI,\AnselmiII,\amax} in the asymptotically free gauge theories 
via the 't Hooft anomaly matching condition \tHooft\ 
by\foot{In this paper, 
we are not interested in the overall normalization of 
the $a$-function and will thus omit it. In order to get the conventionally 
normalized $a$-function, one needs to multiply $3/32$ with the function 
$a(s)$ of this paper.} 
\eqn\afun{
a(s)=\sum_{A\in{\rm UV}}\left[3\left(R_A(s)-1\right)^3-\left(R_A(s)-1\right)
\right]
}
in terms of the $U(1)_R$ charges $R_A(s)$ of the fundamental particles $\phi_A$ 
at high energy. 
The latter condition means that all the eigenvalues of the matrix 
on the left hand side should be negative. Since the matrix is related to 
the two-point functions of the $U(1)$ currents, the unitarity 
requires the latter condition \refs{\amax,\BGISW}. 

However, if there exists the region of the parameter space spanned by 
$\{s^i\}$ where the $U(1)_R$ charge $R_I(s)$ of an operator $\Phi_I$ seems to 
violate the unitarity bound, one needs to subtract the contribution of $\Phi_I$ 
from the function $a(s)$, since it becomes a free field with the fixed 
$U(1)_R$ charge $R_I=2/3$ at the point of the parameter space to maintain 
the unitarity of the theory\foot{Indeed, as discussed in \amax, when $\Phi_I$ 
becomes free, an accidental $U(1)_\Phi$ symmetry appears and enables us to 
fix the $U(1)_R$ charge of $\Phi_I$ to 2/3 via $a$-maximization, while keeping the $U(1)_R$ charge 
of the other operators unchanged.} \refs{\bound}. 
Therefore, in different regions with different gauge invariant operators 
hitting  the unitarity bounds, one needs to improve the a-function $a(s)$ 
and examine the existence of a local maximum in each of the regions. 
Following the prescription of the paper\foot{We will see below that the 
prescription is consistent with the electric-magnetic duality, when the hitting 
operators are elementary fields in the magnetic theory.} \bound, one need to 
modify the a-function $a(s)$ as 
\eqn\Ffun{
a(s)-F(R_I)+F_0, 
\qquad
F(x)=3(x-1)^3-(x-1), 
}
if a field $\Phi_I$ decouples to be free at some points of the parameter 
space $\{s^i\}$. 
Furthermore, since $R_I=2/3$ on the boundary of these two regions, 
the function $a(s)$ and its first derivative with respect to the parameters 
$s^i$ have the same values as $a(s)-F(R_I)+F_0$ and its first derivative, 
respectively. Thus, one obtains a continuous function on the whole parameter 
space $\{s^i\}$, which is not necessarily a third order polynomial in the 
parameters $s^i$ as a whole. This suggests that 
we could find more than one local maximum of the whole function $a(s)$. 

Within a region with the same content of decoupling gauge invariant operators 
%hitting the unitarity bound 
in the whole parameter space $\{s^i\}$, one can find at most 
a single local maximum, but in another region, one could obtain another 
local maximum, where one should find the different content of interacting 
massless gauge invariant operators. It may suggest that one could find more 
than one local maximum over the whole parameter space to lose definitive 
results on which linear combination of the $U(1)$ symmetries is the 
superconformal $U(1)_R$ symmetry. The weaker version of the very recent proposal 
in the paper \Keni\ could however be a way out of this problem. 
It says that ``{\it the correct IR phase is the one with the larger value of 
the conformal anomaly $a$}''. It would thus be very interesting to 
find models with more than one local maximum of the function $a(s)$ and to 
study the renormalization group flow in the models. 

In this paper, we will find a local maximum of the whole function $a(s)$ 
for $7\leq{N}_Q\leq21$. However, we haven't completely confirmed that 
it is a unique local maximum, mainly due to difficulty in identifying 
the massless spectrum in the infrared and also due to the lack of our 
understanding about $a$-maximization applied to a gauge invariant operator 
in a non-trivial representation of the Lorentz group, as we will discuss later. 
We therefore will have to leave this question to the future. 
The problem, even if it really exists, doesn't affect our results on 
the local maximum in this paper, except for the uniqueness of it. 

The identification of the superconformal $U(1)_R$ symmetry enables us 
to know which gauge invariant chiral primary operators hit the unitarity bounds. 
We will discuss the renormalization group flow of the dual pair of our theories 
into another dual pair and the auxiliary field method of the electric theory. 
We will also check the consistency of our results by finding which operators 
as perturbation in the superpotential are irrelevant at the superconformal 
fixed point. Since the operators hitting the unitarity bounds are free, 
the couplings to them in the superpotential go to zero. 
Therefore, the corresponding perturbations must be irrelevant at the infrared 
fixed point. 

To keep this paper within reasonable length, we will report our results for $Spin(10)$ gauge theory with more than one spinor 
in another paper \KOTY. 

This paper is organized as follows: in section two, we will give a brief 
review on the electric-magnetic duality \refs{\PSX,\kawano} in the $Spin(10)$ gauge theory. In section three, the whole $a$-function will be constructed 
and the $U(1)_R$ charges are determined via the $a$-maximization method. 
Section four is devoted to summary and discussion, where 
the flow of our dual pair into another dual pair and the above auxiliary field 
method will be discussed.

%%%%%%%%%%%%%%%%%%%%%%%%%%%%%%%%%%%%%%%%%%%%%%%%%%%%%%%%%%%%%%%%%%%%%%%%%%%%%
\newsec{The Electric-Magnetic Duality in the $Spin(10)$ Gauge Theories}
%%%%%%%%%%%%%%%%%%%%%%%%%%%%%%%%%%%%%%%%%%%%%%%%%%%%%%%%%%%%%%%%%%%%%%%%%%%%%

Let us begin with a brief review of the electric-magnetic duality 
\refs{\PSX,\kawano} of the theory we will consider in this paper. 
We study four-dimensional ${\cal N}=1$ supersymmetric $Spin(10)$ gauge theory 
with a single spinor $\Psi$ and $N_Q$ vectors $Q^i$ ($i=1,\cdots,N_Q$) 
and with no superpotential. Under the non-anomalous global symmetries 
$SU(N_Q)\times{U(1)}\times{U(1)_\lambda}$, the matter superfields $\Psi$ and 
$Q^i$ transform as $(1,N_Q,0)$ and $(N_Q,-2,{N_Q-6\over{N_Q}})$, respectively, 
as our convention. 

It is believed \refs{\PSX,\kawano} that it is in the 
non-Abelian Coulomb phase for $7\leq{N}_Q\leq21$, where it is asymptotically 
free and has the dual or magnetic description. 
The magnetic theory is given by $SU(N_Q-5)$ gauge theory with 
$N_Q$ antifundamentals $\bar{q}_i$, a single fundamental $q$, 
a symmetric tensor $s$ and singlets $M^{ij}$ and $Y^i$ with the superpotential 
\eqn\Wmag{
W_{\rm mag}={\tilde{h}\over\tilde\mu^2}M^{ij}\bar{q}_i\,s\,\bar{q}_j
+{\tilde{h}'\over\tilde\mu^2}Y^i\,q\bar{q}_i+{1\over\tilde\mu^{N_Q-8}}\det{s}.
}
Only for $N_Q=7$, one have the additional term 
$$
{\tilde{h}''\over\tilde\mu^{15}}\epsilon_{i_{1}\cdots{i}_{7}}
\epsilon_{j_{1}\cdots{j}_{7}}
M^{i_{1}j_{1}}\cdots{}M^{i_{6}j_{6}}Y^{i_{7}}Y^{j_{7}}
$$
in the above superpotential $W_\mag$, as discussed in \refs{\PSX,\kawano}. 

Thus, at the infrared fixed point, it is an ${\cal N}=1$ superconformal field 
theory with the superconformal $U(1)_R$ symmetry out of linear 
combinations of the ${U(1)}\times{U(1)_\lambda}$ symmetries. 
The $U(1)_R$ charges of the matter fields should satisfy 
the Adler-Bell-Jackiw $U(1)$ anomaly cancellation condition 
\refs{\Adler,\BellJackiw}
$$
2R(\Psi)=-N_QR(Q)+(N_Q-6), 
$$
and thus turn out to be given by 
$
R(Q)={N_Q-6\over{N}_Q}-2x, R(\Psi)=N_Q\,x, 
$
respectively, for a fixed real number $x$.  
According to $a$-maximization \amax, the fixed real number $x$ or 
equivalently the $U(1)_R$ charge $R\equiv{R}(Q)$ of the vectors $Q^i$ must 
be the solution to the conditions \amaximization. 
If there aren't any gauge invariant operators hitting the unitarity bounds, 
one can invoke the 't Hooft anomaly matching condition \tHooft\ 
to give the $a$-function in terms of the elementary fields as
\eqn\aUVfun{
a_0(R)=90+16 F\left[R\left(\Psi\right)\right] 
+10N_Q F\left[R\left(Q\right)\right],
}
where the function $F(x)$ was defined in \Ffun. The first term on the right hand 
side comes from the contribution of the gaugino, which are forty-five Weyl 
spinors of charge one with respect to the $U(1)_R$ symmetry, thus giving 
$45\times\left[3R(\lambda)^3-R(\lambda)\right]=90$.

In this theory, there are several gauge invariant chiral operators\foot{
We omit for simplicity explicit indices of the gauge group $Spin(10)$, 
which we assume are obvious in context. In particular, the gauge invariant 
operators are built out of the elementary chiral superfields with 
the invariant Kronecker deltas and the invariant tenth 
rank antisymmetric tensors.}
$$\eqalign{
&M^{ij}=Q^iQ^j, 
\qquad
Y^i=Q^i\,\Psi\Gamma^{(1)}\Psi, 
\qquad
B^{{i}_1\cdots{i}_5}=Q^{i_1}\cdots{Q}^{i_{5}}\,\Psi\Gamma^{(5)}\Psi,
\cr
&D_{0}^{{i}_1\cdots{i}_6}=Q^{i_1}\cdots{Q}^{i_{6}}\,W_{\alpha}W^{\alpha},
\quad
D_{1\,\alpha}^{{i}_1\cdots{i}_8}=Q^{i_1}\cdots{Q}^{i_{8}}\,W_{\alpha},
\quad 
D_{2}^{{i}_1\cdots{i}_{10}}=Q^{i_1}\cdots{Q}^{i_{10}},
\cr
&E_{0}^{{i}_1\cdots{i}_5}
=Q^{i_1}\cdots{Q}^{i_{5}}\,\Psi\Gamma^{(1)}\Psi\,W_{\alpha}W^{\alpha},
\quad 
E_{1\,\alpha}^{{i}_1\cdots{i}_7}
=Q^{i_1}\cdots{Q}^{i_{7}}\,\Psi\Gamma^{(1)}\Psi\,W_{\alpha},
\cr
&E_{2}^{{i}_1\cdots{i}_{9}}=Q^{i_1}\cdots{Q}^{i_{9}}\,\Psi\Gamma^{(1)}\Psi,
}$$
where $\Psi\Gamma^{(1)}\Psi$ and $\Psi\Gamma^{(5)}\Psi$ are bilinear 
combinations out of the spinor $\Psi$ together with the gamma matrices 
of the gauge group $Spin(10)$ and transform in the vector representation 
and as a fifth rank antisymmetric tensor, respectively. The chiral superfield 
$W_{\alpha}$ is the $Spin(10)$ gauge superspace field strength in the adjoint 
representation of the gauge group. Note that depending upon 
the number $N_Q$ of the vectors, some of the gauge invariant operators aren't 
available, as illustrated in Figure 1. 
%%%%%%%%%%%%%%%%%%%%%%%%%%%%%%%%%%%%%%%%%%%%%%%%
\fig{\it The number $N_Q$ of the vectors $Q^i$ where the gauge invariant operators 
$D_n$ and $E_n$ exist.
}{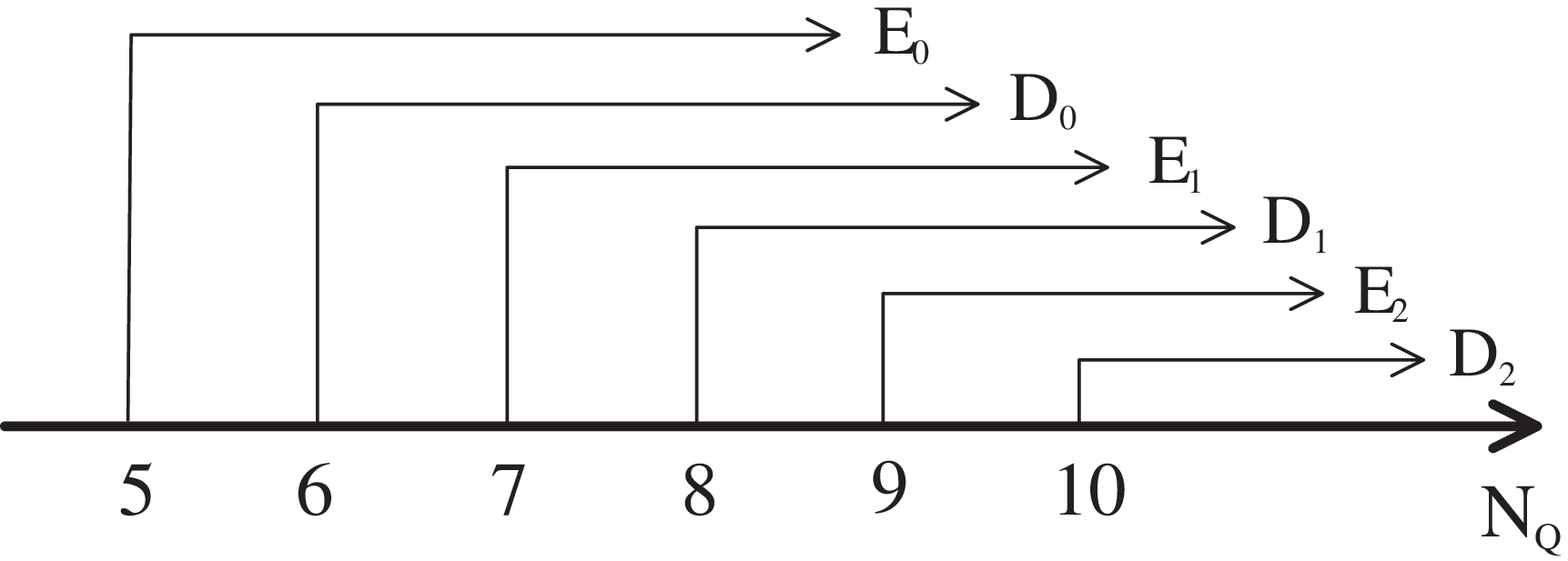}{8.5 truecm}
\figlabel\zuI
%%%%%%%%%%%%%%%%%%%%%%%%%%%%%%%%%%%%%%%%%%%%%%%%
In the magnetic theory, the gauge invariant fields $M^{ij}$ and 
$Y^i$ are introduced as elementary fields, while the other gauge invariant 
operators can be constructed out of the antifundamentals $\bar{q}_i$, 
the fundamental $q$, the symmetric tensor $s$, and the dual gauge superspace 
field strength $\widetilde{W}_{\alpha}$ as 
$$\eqalign{
&\left(*B\right)_{j_1\cdots{}_{N_Q-5}}\sim
\bar{q}_{j_1}\cdots\bar{q}_{j_{N_Q-5}},
\cr
&\left(*D_0\right)_{j_1\cdots{j}_{N_Q-6}}\sim
q\left(s\bar{q}_{j_1}\right)\cdots\left(s\bar{q}_{j_{N_Q-6}}\right),
\quad
\left(*D_{1\alpha}\right)_{j_1\cdots{j}_{N_Q-8}}\sim
q\,\left(s\bar{q}_{j_1}\right)\cdots
\left(s\bar{q}_{j_{N_Q-8}}\right)\left(\widetilde{W}_{\alpha}s\right),
\cr
&\left(*D_{2}\right)_{j_1\cdots{j}_{N_Q-10}}\sim
q\,\left(s\bar{q}_{j_1}\right)\cdots
\left(s\bar{q}_{j_{N_Q-10}}\right)\left(\widetilde{W}_{\alpha}s\right)
\left(\widetilde{W}^{\alpha}s\right),
\cr
&\left(*E_0\right)_{j_1\cdots{j}_{N_Q-5}}\sim
\left(s\bar{q}_{j_1}\right)\cdots\left(s\bar{q}_{j_{N_Q-5}}\right),
\quad
\left(*E_{1\alpha}\right)_{j_1\cdots{j}_{N_Q-7}}\sim
\left(s\bar{q}_{j_1}\right)\cdots
\left(s\bar{q}_{j_{N_Q-7}}\right)\left(\widetilde{W}_{\alpha}s\right),
\cr
&\left(*E_{2}\right)_{j_1\cdots{j}_{N_Q-9}}\sim
\left(s\bar{q}_{j_1}\right)\cdots
\left(s\bar{q}_{j_{N_Q-9}}\right)\left(\widetilde{W}_{\alpha}s\right)
\left(\widetilde{W}^{\alpha}s\right),
}$$
where the operation $*$ on the gauge invariant operators denotes the Hodge 
duality with respect to the flavor $SU(N_Q)$ symmetry. 
It is interesting to note that the classical moduli $D_2$ and $E_2$ 
in the electric theory are given by the gauge invariant operators 
containing the dual gaugino superfield\foot{The 
operators $D_0$ and $E_0$ can be rewritten as $(q\bar{q})\cdot{B}\cdot{M}$ 
and $\det{s}\cdot{B}$, respectively in the magnetic theory. The operators 
$q\bar{q}_j$ and $\det{s}$ are redundant, as will be seen below. Therefore, 
they should be redundant at least for $7\leq N_Q\leq 21$.} 
$\widetilde{W}_{\alpha}$. Besides the above operators, in the magnetic theory, 
there are other gauge invariant operators such as 
$N_{ij}=\bar{q}_i\,s\,\bar{q}_j$, $\det{s}$, $q\bar{q}_i$. 
They, however, are redundant, due to the F-term condition from 
the superpotential $W_{\rm mag}$. 

\topinsert
\parasize=1in

\begintable
Gauge Invariant Operators ${\cal O}$ \|  the $U(1)$ charge 
\| the $U(1)_R$ charge $R({\cal O})$  \cr
$M{\sim}Q^2$ \|  $-4$  \| $2R$         \nr
$Y{\sim}Q\Psi^2$ \| $2N_{Q}-2$  \| $N_Q-6-(N_Q-1)R$        \nr
$B{\sim}\,Q^5\Psi^2{\sim}\,\bar{q}^{N_Q-5}$\|$2N_Q-10$\|$N_Q-6-(N_Q-5)R$\nr
$D_n{\sim}Q^{6+2n}W^{2-n}{\sim}\,\bar{q}^{N_Q-6-2n}s^{N_Q-6-n}\widetilde{W}^nq$ 
\|  $-4n-12$ \| $(2n+6)R-(n-2)$ \nr
$E_n{\sim}Q^{5+2n}\Psi^2W^{2-n}{\sim}\,\bar{q}^{N_Q-5-2n}s^{N_Q-5-n}
\widetilde{W}^n$ \| $-4n+2N_{Q}-10$ \| $(2n-N_Q+5)R-(n-N_Q+4)$
\endtable
\nobreak 
\centerline{\it Table 1: The charges of the gauge invariant 
operators with respect to the $U(1)\times{U}(1)_R$ symmetry.}
\endinsert

It follows from the superpotential $W_{\rm mag}$ that 
the elementary fields $\bar{q}_i$, $q$, and $s$ 
in the magnetic theory have the charges 
$\left(2,{N_Q-6\over{N}_Q-5}-R\right)$, 
$\left(-2N_Q,N_Q(R-1)+{7N_Q-34\over{N}_Q-5}\right)$, 
%$\left(-2N_Q,N_QR-(N_Q-8)-{N_Q-6\over{N}_Q-5}\right)$, 
and $\left(0,{2\over{N}_Q-5}\right)$, 
respectively, with respect to the $U(1)\times{U(1)}_R$ symmetry. 
These charges satisfy the Adler-Bell-Jackiw $U(1)$ anomaly cancellation 
condition \refs{\Adler,\BellJackiw} 
and further are consistent with the mapping of the gauge 
invariant operators between the electric side and the magnetic one as shown in Table 1. 

For later convenience, we introduce another dual pair, into which we will 
see below that the previous dual pair flows in the infrared for $7\le{N_Q}\le9$. 
When in the electric theory, turning on the superpotential 
$$
W_{\ele}={1\over\mu^2}N_{ij}Q^iQ^j,
$$
one finds that the moduli $M^{ij}$ are eliminated because of the F-term 
condition 
$$
{\partial\ \over \partial N_{ij}}W_{\ele}={1\over\mu^2}Q^iQ^j=0, 
$$
and instead that the new moduli $N_{ij}$ show up. 
It can be seen in the magnetic theory that 
the first term $M^{ij}\bar{q}_i\,s\,\bar{q}_j$ 
in the superpotential $W_{\rm mag}$ has to be turned off to decouple the gauge 
singlets $M^{ij}$. The new F-term condition from the magnetic superpotential 
doesn't impose any constraints on the gauge invariant operator 
$N_{ij}=\bar{q}_i\,s\,\bar{q}_j$. Although the use of $N_{ij}$ seems 
the abuse of the notation, the two on the both sides are in the same 
representation 
 $$
N_{ij} : \left(\symbar,4,2-R(M)=2(1-R)\right)
$$
of the global symmetries $SU(N_Q)\times{U}(1)\times{U(1)_R}$ 
and can thus be identified. The field $N_{ij}$ will play an important role, 
when the gauge invariant operator $M^{ij}$ hits the unitarity bound in the 
original $Spin(10)$ theory. 
The electric $Spin(10)$ theory with the superpotential $W_{\ele}$ therefore is dual 
to the previous magnetic theory in the absence of the first term in 
the superpotential $W_{\mag}$ and without the singlets $M^{ij}$. 
It is important to note that all the gauge invariant operators discussed just 
above are retained except for $M^{ij}$ even in this dual pair.

%%%%%%%%%%%%%%%%%%%%%%%%%%%%%%%%%%%%%%%%%%%%%%%%%%%%%%%%%%%%%%%%%%%%%%%%%%%%%%
%%%%%%%%%%%%%%%%%%%%%%%%%%%%%%%%%%%%%%%%%%%%%%%%%%%%%%%%%%%%%%%%%%%%%%%%%%%%%%
\newsec{$a$-Maximization in the $Spin(10)$ Theories}

When the gauge invariant chiral primary operators hit the unitarity bounds, 
they decouple from the remaining system as free fields of the $U(1)_R$ charge 
$2/3$. Therefore, following the prescription of the paper \bound, 
one needs to improve the previous $a$-function $a_0(R)$ as 
$$
a(R)=90+16 F\left[R\left(\Psi\right)\right] 
+10N_Q F\left[R\left(Q\right)\right]-\sum_i\left[F[R(\Phi_i)]-F_0\right],
$$
where $\Phi_i$ are the gauge invariant operators hitting the unitarity 
bounds. Conversely, in order to find the solution to the $a$-maximization 
condition \amaximization, one needs to look at all real values of $x$ or 
equivalently $R$. As can be read from Table 1, the following unitarity 
bounds\foot{The unitarity bound for a spin one-half field is given \Mack\ by 
$D\geq{3\over2}$, which gives the bound for $U(1)_R$ charge; $R\geq1$.}
\eqn\Rbound{\eqalign{
&R(M)=2R\ge{2\over3} \qquad \qquad \qquad \qquad \, \ \quad\Rightarrow\quad R\ge{1\over3},
\cr
&R(Y)=N_Q-6-(N_Q-1)R\ge{2\over3}\quad \quad\Rightarrow\quad 
R\le{1\over{N_Q-1}}\left(N_Q-{20\over3}\right),
\cr
&R(B)=N_Q-6-(N_Q-5)R\ge{2\over3}\quad \quad\Rightarrow\quad 
R\le{1\over{N_Q-5}}\left(N_Q-{20\over3}\right),
\cr
&R(D_2)=10R \ge{2\over3} \qquad \qquad \qquad \quad \qquad\Rightarrow\quad 
R\ge{1\over15},
\cr
&R(E_2)=(N_Q-6)-(N_Q-9)R \ge{2\over3} \quad\Rightarrow\quad 
R\le{1\over N_Q-9}\left(N_Q-{20\over3}\right),
\cr
&R(D_1)=8R+1 \ge{1} \qquad \qquad \qquad \, \qquad\Rightarrow\quad R\ge0,
\cr
&R(E_1)=(N_Q-5)-(N_Q-7)R \ge{1} \, \quad\Rightarrow\quad 
R\le{N_Q-6 \over N_Q-7},
}}
divide all real values of $R$ into several regions\foot{We don't take account of 
the unitarity bounds for the gauge invariant operators $D_0$ and $E_0$, since 
they are redundant, as discussed previously. }, as sketched for 
$N_Q=7$ in Figure 2. Note that, when the denominator on the right 
hand side in the conditions for $R$ from the unitarity bounds for $R(E_n)$ 
is zero, the corresponding unitarity bounds are independent of $R$ and are 
always satisfied. 
%
%%%%%%%%%%%%%%%%%%%%%%%%%%%%%%%%%%%%%%%%%%%%%%%%
\fig{\it A sketch of operators hitting the unitarity bounds for the theory with 
$7$ vectors. Each of the regions from $I$ to $IV$ are separated at 
$R(Q)= 1/18$, $1/6$, $1/3$, respectively. The arrows show the regions where 
the corresponding operators hit the unitarity bounds. 
}{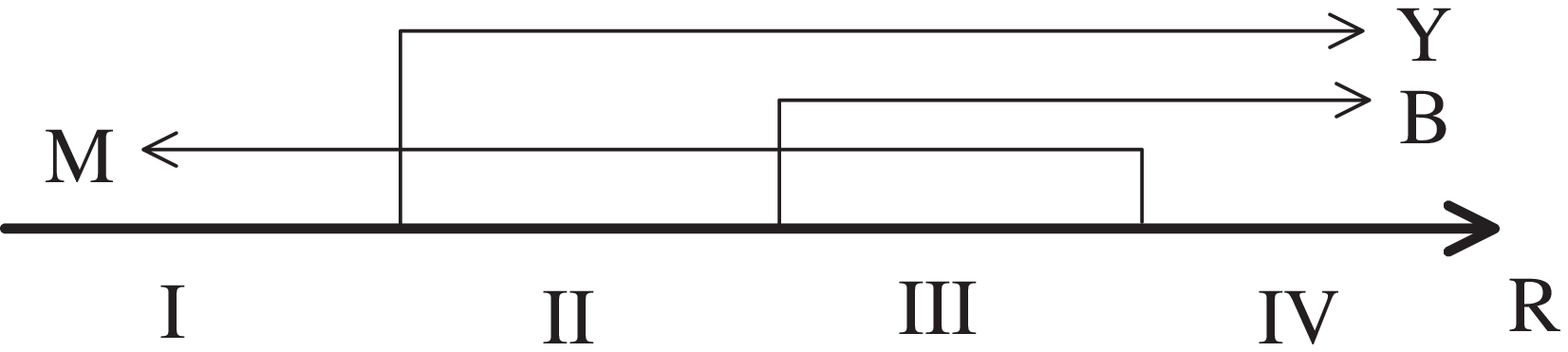}{10 truecm}
\figlabel\zuII
%%%%%%%%%%%%%%%%%%%%%%%%%%%%%%%%%%%%%%%%%%%%%%%%

There is a subtle point about the massless spectrum of the gauge invariant 
operators. The gauge invariant chiral superfields $M$, $Y$, $B$, $D_2$, 
and $E_2$ parametrize the classical moduli space of the electric theory, 
and it is thus natural to consider that they are in the massless spectrum. 
The magnetic theory implies that the operators $D_0$ and $E_0$ are redundant 
by the F-term condition, as discussed before. However, 
there seems no compelling arguments about whether the Lorentz spinor operators 
$D_1$ and $E_1$ are massless or massive. 
If they were massless and hit the unitarity bounds, one would encounter a 
problem in the calculation of a trial $a$-function. Since they are in the 
spinor representation of the Lorentz group, at present we don't understand how 
the $a$-maximization method can be extended for those operators.

In fact, in some regions of the space of the $U(1)_R$ charge, the spinor 
operators appear to violate the unitarity bounds. 
Therefore, in that case, we have to assume 
that they are massive and don't contribute to the $a$-function in the infrared. 
We henceforth will not take account of the operators $D_1$ and $E_1$ 
upon the use of the $a$-maximization method. However, if they are actually 
massless, our results could not be correct in the regions where they hit 
the unitarity bounds given above. We will see below that the solutions to the 
$a$-maximization condition \amaximization\ 
are found in the other region, where $D_1$ and $E_1$ don't hit the bounds. 
Therefore, the solutions remain valid, even when they are massless. 

We will demonstrate in detail the $a$-maximization procedure for 
the case of $N_Q=7$ vectors $Q^i$, and then will report our results 
on the other value of $N_Q$. Before proceeding, let us make a 
comment on the structure of divided regions of the space of the $U(1)_R$ charge 
$R$. When one looks at the operators hitting the unitarity bounds from minus 
large value of $R$ to positive large value, the order of the hitting operators 
on the line of $R$ may change, as one change the number $N_Q$. It turns out 
from the unitarity bounds \Rbound\ that, 
although one needs to consider each case for $N_Q=7,8,9$, one can study 
the other cases $N_Q\geq10$ in a unified manner, because the order of hitting 
of the operators remains unchanged for the latter cases. 
One also finds that, for the latter cases, there is the region where none of 
the gauge invariant operators hit the unitarity bounds, but 
no such regions for the former cases of $N_Q=7,8,9$. 

\topinsert
\parasize=1in

\begintable
     \| Hitting Operators \|  Hitting Regions   \cr
$I$ \| $M$ \| $R\leq{1\over18}$ \nr 
$II$ \| $M$, $Y$ \| ${1\over18}\leq{R}\leq{1\over6}$ \nr
$III$ \| $M$, $Y$, $B$ \| ${1\over6}\leq{R}\leq{1\over3}$ \nr
$IV$ \| $Y$, $B$ \| ${R}\geq{1\over3}$
\endtable
\centerline{\it Table 2: The four regions of the $U(1)_R$ charge $R$ 
for $N_Q=7$.}
\endinsert

In the case of $N_Q=7$ vectors, there are four regions dividing the space 
of the $U(1)_R$ charge $R$, as can be seen in Table 2 and as illustrated 
in Figure 2. In each region, one finds the above $a$-function $a(R)$ as 
$$\eqalign{
&a(R)=a_0(R)-{N_Q(N_Q+1)\over2}\left(F[R(M)]-F_0\right),
%\quad {\rm for}~R\leq{1\over18},
\quad \left(R\leq{1\over18}\right),
\cr
&a(R)=a_0(R)-{N_Q(N_Q+1)\over2}\left(F[R(M)]-F_0\right)
-N_Q\left(F[R(Y)]-F_0\right),
%\quad {\rm for}~{1\over18}\leq{R}\leq{1\over6},
\quad \left({1\over18}\leq{R}\leq{1\over6}\right),
\cr
&a(R)=a_0(R)-{N_Q(N_Q+1)\over2}\left(F[R(M)]-F_0\right)
-N_Q\left(F[R(Y)]-F_0\right)
\cr
&\hskip 6.5cm 
-{N_Q!\over(N_Q-5)!5!}\left(F[R(B)]-F_0\right),
%{\rm for}~{1\over6}\leq{R}\leq{1\over3},
\quad \left({1\over6}\leq{R}\leq{1\over3}\right),
\cr
&a(R)=a_0(R)-N_Q\left(F[R(Y)]-F_0\right)
-{N_Q!\over(N_Q-5)!5!}\left(F[R(B)]-F_0\right),
%\quad {\rm for}~{R}\geq{1\over3},
\quad \left({R}\geq{1\over3}\right),
}$$
with $N_Q=7$ substituted. The whole function $a(R)$ is illustrated in Figure 3, 
which explicitly shows that it isn't a third order polynomial of $R$, 
but gives two local minima. As can be seen in Figure 3, there is a unique 
local maximum, where only the mesons $M^{ij}$ are free and the $U(1)_R$ 
charge gives $R={1/30}$. It is the local maximum 
\eqn\asol{
R={3N_Q^2 - 21N_Q - 12 + 2\sqrt{-(N_Q-6)(N_Q^2-29N_Q+73)}\over 3(N_Q+3)(N_Q-1)}
}
of the function $a_0(R)-(N_Q(N_Q+1)/2)F(2R)$ for $N_Q=7$. 
%%%%%%%%%%%%%%%%%%%%%%%%%%%%%%%%%%%%%%%%%%%%%%%
\fig{\it The whole a-function $a(R)$ for $N_Q=7$.
}{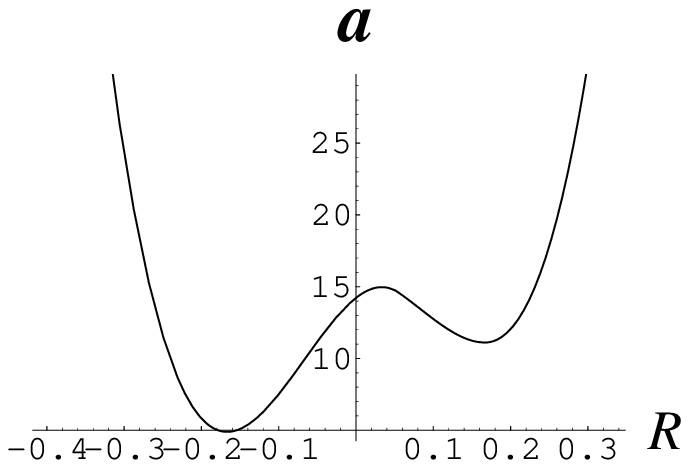}{6.8 truecm}
\figlabel\zuIII
%%%%%%%%%%%%%%%%%%%%%%%%%%%%%%%%%%%%%%%%%%%%%%%%

%
\topinsert
\parasize=1in

\begintable
     \| Hitting Operators \|  Hitting Regions   \cr
$I+II$ \| $M$ \| $R\leq{1\over N_Q-1}(N_Q-{20\over3})$ \nr 
$III$ \| $M$, $Y$ \| ${1\over N_Q-1}(N_Q-{20\over3})\leq{R}\leq{1\over3}$ \nr
$IV$ \| $Y$ \| ${1\over3}\leq{R}\leq{1\over N_Q-5}(N_Q-{20\over3})$ \nr
$V+VI$ \| $Y$, $B$ \| ${R}\geq{1\over N_Q-5}(N_Q-{20\over3})$
\endtable
\centerline{\it Table 3: The four regions of the $U(1)_R$ charge $R$ 
for $N_Q=8,9$.}
\endinsert

For $N_Q=8,9$, as can be seen in Table 3, there are four regions 
on the line of the $U(1)_R$ charge $R$, as in Figure 4. 
As is different from the case of 
$N_Q=7$, there is no region where the three gauge invariant operators 
$M$, $Y$, and $B$ hit the unitarity bounds at the same time, but 
appears a new region $IV$, where only the operator $Y^i$ hits the bound. 
Only for $N_Q=9$, the exotic $E_2$ is available, but it doesn't violate 
the unitarity bound over all the values of $R$. If the spinor exotics $E_1$ 
and $D_1$ were massless, our results for the regions $I$ and $VI$ would be 
incomplete. The whole $a$-functions $a(R)$ are similar to the one for 
$N_Q=7$ and has, in the region $II$, a single local maximum given by \asol\ 
with $N_Q=8,9$ substituted for each case, 
where also only the meson $M^{ij}$ is hitting the unitarity bound 
to be free in the infrared. Note that the local maximum would be retained 
even after taking account of the exotics $E_1$ and $D_1$. 

%%%%%%%%%%%%%%%%%%%%%%%%%%%%%%%%%%%%%%%%%%%%%%%%
\fig{\it The operators hitting the unitarity bounds for the theory with $N_Q=8$ 
and $9$ vectors. The arrows show the regions where the corresponding 
operators hit the unitarity bounds.
}{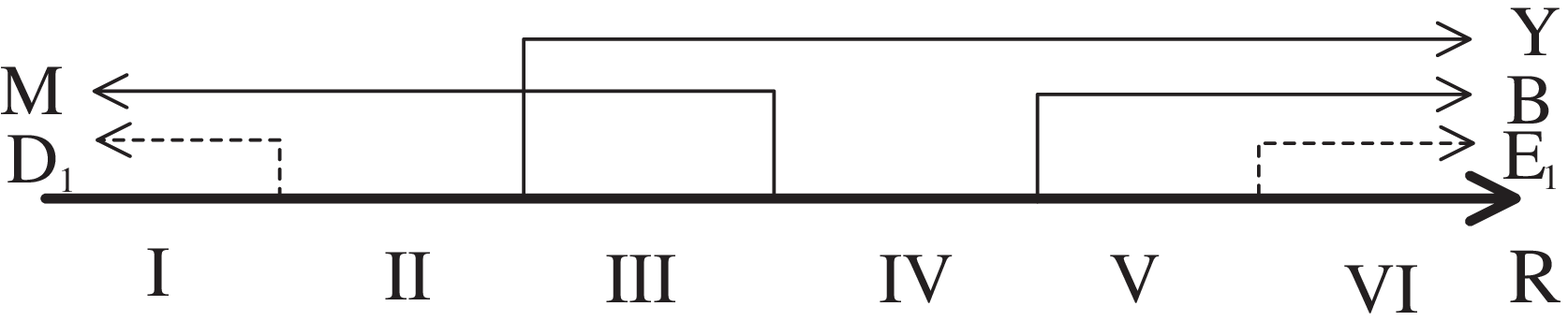}{10 truecm}
\figlabel\zuIV
%%%%%%%%%%%%%%%%%%%%%%%%%%%%%%%%%%%%%%%%%%%%%%%%

%
\topinsert
\parasize=1in

\begintable
     \| Hitting Operators \|  Hitting Regions   \cr
$I+II$ \| $M$, $D_2$ \| $R\leq{1\over15}$ \nr 
$III$ \| $M$  \| ${1\over15}\leq{R}\leq{1\over3}$ \nr
$IV$ \| no operators \| ${1\over3}\leq{R}\leq{1\over N_Q-1}(N_Q-{20\over3})$ \nr
$V$ \| $Y$ \| ${1\over N_Q-1}(N_Q-{20\over3})\leq{R}\leq{1\over N_Q-5}(N_Q-{20\over3})$ \nr
$VI+VII$ \| $Y$, $B$ \| ${1\over N_Q-5}(N_Q-{20\over3})\leq{R}\leq{1\over N_Q-9}(N_Q-{20\over3})$ \nr
$VIII$ \| $Y$, $B$, $E_2$ \| ${R}\geq{1\over N_Q-9}(N_Q-{20\over3})$
\endtable
\nobreak
\centerline{\it Table 4: The six regions on the line of the $U(1)_R$ charge $R$ 
for $10\leq{N_Q}\leq21$.}
\endinsert

For $10\leq{N_Q}\leq21$, it is remarkable that there exists the region $IV$ 
with no gauge invariant operators hitting the unitarity bounds, as shown in 
Figure 5. The parameter space of the $U(1)_R$ charge $R$ is divided into 
six regions, as can be seen in Table 4. 
The regions $I$, $VII$, and $VIII$ could be incomplete due to the exotics 
$D_1$ and $E_1$. The whole $a$-function $a(R)$ has a similar shape to the 
one in Figure 3. One finds a unique local maximum at 
$$
R={3N_Q^2-24N_Q-15+\sqrt{2885-N_Q^2}\over3(N_Q^2-5)}
$$
in the regions $IV$, where no operator hits the unitarity bound. 
The local maximum also remains valid even after taking account of the 
unitarity bounds of $D_1$ and $E_1$. 
%%%%%%%%%%%%%%%%%%%%%%%%%%%%%%%%%%%%%%%%%%%%%%%%
\fig{\it The operators hitting the unitarity bounds for the theory with 
$10\leq N_Q\leq 21$ vectors. The arrows show the regions where the 
corresponding operators hit the unitarity bounds.
}{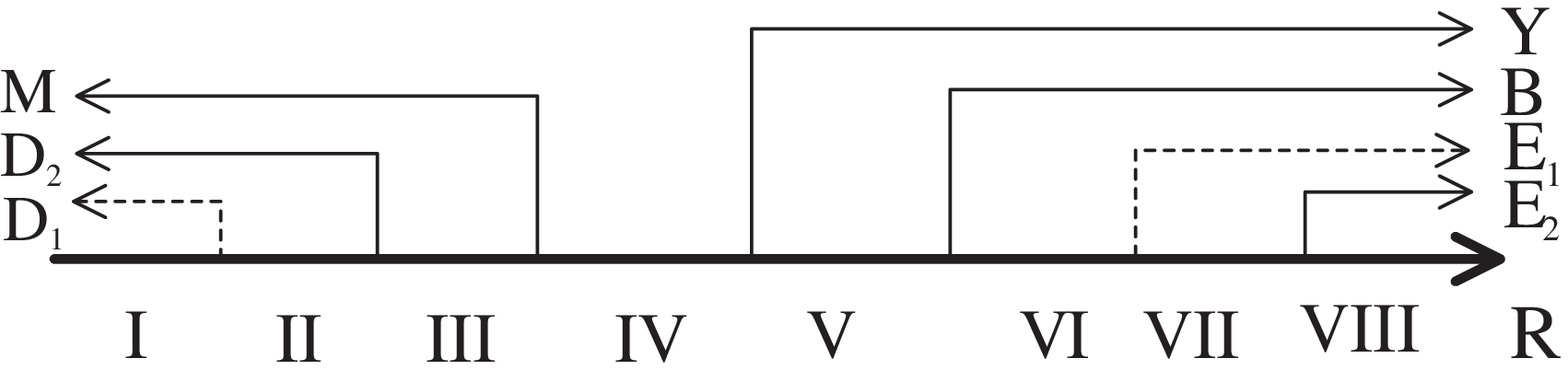}{10 truecm}
\figlabel\zuV
%%%%%%%%%%%%%%%%%%%%%%%%%%%%%%%%%%%%%%%%%%%%%%%%

Finally, let us make a brief comment on the weaker version of 
the $a$-theorem \Cardy. 
From the above results, one can immediately calculate the $a$-function 
$a_{{\rm IR}}$ in the infrared and compare it to the one $a_{{\rm UV}}$ 
at high energy, as in Figure 6. The $a$-function $a_{{\rm UV}}$ counts 
the number of the fundamental fields as free fields, each of which 
contribute $F_0$ to it. One can see that the inequality 
$$
a_{{\rm IR}}<a_{{\rm UV}}
$$
actually holds for $7\leq{N}_Q\leq21$. 
%%%%%%%%%%%%%%%%%%%%%%%%%%%%%%%%%%%%%%%%%%%%%%%%
\fig{\it The graph on the left shows the $R$ charges $R(Q)$ (solid line) and $R(\Psi)$ (dotted line). The graph on the right depicts the central charges at the UV free fixed (dotted line) and IR fixed point(solid line).
}{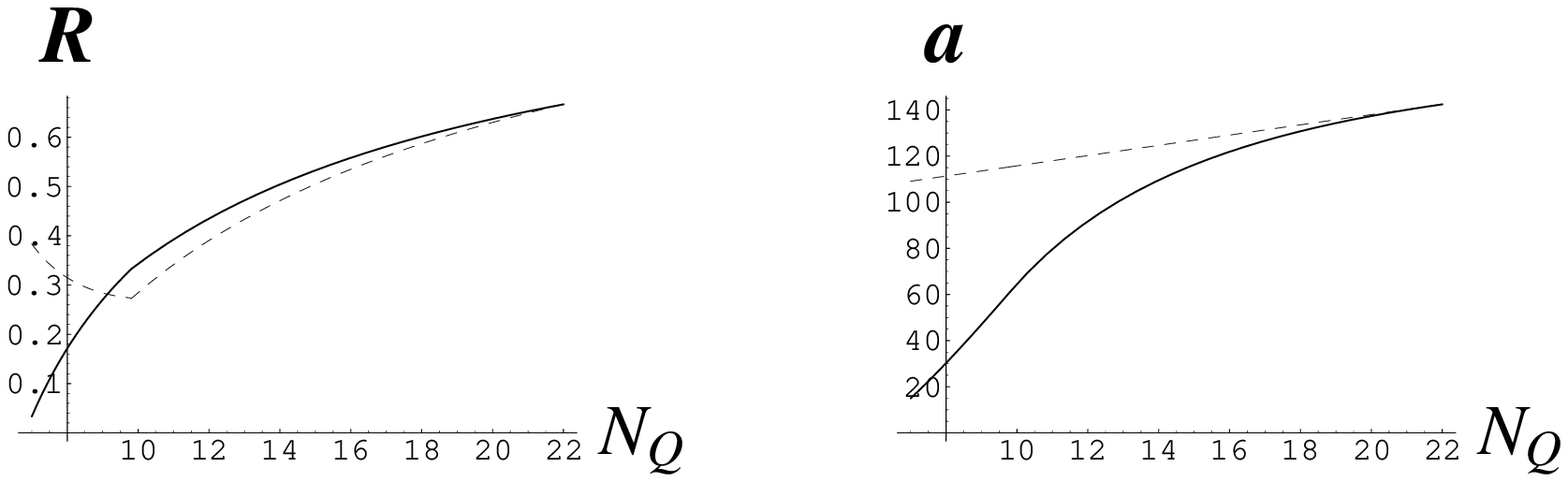}{11 truecm}
\figlabel\zuV
%%%%%%%%%%%%%%%%%%%%%%%%%%%%%%%%%%%%%%%%%%%%%%%%

%%%%%%%%%%%%%%%%%%%%%%%%%%%%%%%%%%%%%%%%%%%%%%%%%%%%%%%%%%%%%%%%%%%%%%%%%%

\newsec{Summary and Discussion}

We have found a unique local maximum of the $a$-function for $10\leq{N_Q}\leq21$, where there are no gauge invariant chiral primary operators 
hitting the unitarity bounds. On the other hand, for $7\leq{N_Q}\leq9$, 
one can also obtain a unique local maximum of the $a$-function, but 
at the local maximum, one finds that the gauge invariant operators $M^{ij}$ 
are free fields at the infrared fixed point, otherwise it would violate 
the unitarity of the theory. The existence of the local maximum is 
consistent with the conjecture \refs{\PSX,\kawano} that this theory is 
in the non-Abelian Coulomb phase for $7\leq{N_Q}\leq21$. As discussed previously, the gauge singlet spinors $D_1$ and $E_1$ could invalidate the uniqueness of 
the local maximum in the above mentioned regions of $R$. It would be 
interesting if we could extend the $a$-maximization method for such 
Lorentz spinor operators. 

One might wonder whether the same results 
could be obtained in the magnetic $SU(N_Q-5)$ theory. However, this 
is automatically guaranteed by the electric-magnetic duality. In fact, 
since the magnetic theory saturates the anomalies of all 
the global symmetries of the electric theory \refs{\PSX,\kawano}, \ie 
the duality satisfies the 't Hooft anomaly matching condition \tHooft,  
and since the both theories give the same gauge invariant operators \PSX, 
the $a$-function in the magnetic theory is identical to the one in the 
electric theory, even when the gauge invariant operators hit the unitarity bounds.

We have found so far that the mesons $M^{ij}$ are free in the deepest 
infrared for $7\leq{N_Q}\leq9$. As discussed for the $Spin(7)$ theory 
in Introduction, it implies that 
the parameter $\tilde{h}$ in the superpotential $W_\mag$ of the magnetic theory 
goes to zero, as one goes to the infrared, as can be seen from \Wmag. 
Since the equations 
of motion gives
$$
{\partial\ \over\partial M^{ij}}W_\mag={\tilde{h}\over\tilde\mu^2}N_{ij}=0, 
$$
where $N_{ij}=\bar{q}_i\,s\,\bar{q}_j$, if $\tilde{h}$ weren't zero, 
the gauge invariant operators $N_{ij}$ would be redundant. 
This is indeed the case for $10\leq{N}_Q\le21$. However, for $7\leq{N_Q}\leq9$, 
since $\tilde{h}$ goes to zero\foot{For $N_Q=7$, the parameter $\tilde{h}''$ also goes to zero.}, the operators $N_{ij}$ don't have to be 
redundant. Therefore, $N_{ij}$ should be new degrees of freedom. 
In fact, at the infrared fixed point, 
the term $M^{ij}\bar{q}_i\,s\,\bar{q}_j$ can be turned on as a perturbation 
in the superpotential $W_\mag$ with vanishing $\tilde{h}$. At the point, 
the $U(1)_R$ charge of the perturbation $M^{ij}\bar{q}_i\,s\,\bar{q}_j$ exceed 
two, since the $U(1)_R$ charges of $M^{ij}$ and $\bar{q}_i\,s\,\bar{q}_j$ are 
equal to $2/3$ and $2-2R(Q)\geq4/3$, respectively. 
Therefore, we can see that it is 
an irrelevant operator at the infrared fixed point. 
This is consistent with the fact that the parameter $\tilde{h}$ goes to zero 
in the infrared. 

Furthermore, one can easily see that the magnetic theory with vanishing 
$\tilde{h}$ in the superpotential \Wmag\ is dual to the same $Spin(10)$ theory 
but with the superpotential
$$
W_\ele=N_{ij}Q^iQ^j, 
$$
with the gauge singlets $N_{ij}$ and the free singlets $M^{ij}$. 
The singlets $N_{ij}$ can be 
identified with $\bar{q}_i\,s\,\bar{q}_j$. Therefore, the magnetic theory 
of the original dual pair flows into the magnetic one of another dual pair 
at the infrared fixed point. It suggests that the original electric theory 
flows into the electric theory with the superpotential $W_\ele$. 

In the electric theory with the superpotential $W_\ele$, we can perform 
the $a$-maximization procedure in a similar way to what we have done in this 
paper. The region where no gauge invariant operators hit 
the unitarity bounds is identical on the line of the $U(1)_R$ charge $R$ 
to the region where only the operators $M^{ij}$ hit the unitarity bound 
in the original electric theory. In the region, the trial $a$-function 
can be calculated in terms of the fundamental fields at high energy 
in the former theory to give 
$$
a_0(R)+{N_Q(N_Q+1)\over2}F[R(N)]+{N_Q(N_Q+1)\over2}F_0,
$$
where $a_0(R)$ is given in \aUVfun, and $F_0$ is the contribution from 
the free singlets $M^{ij}$. Since the function $F(x)$ satisfies the relation 
\eqn\Fmassive{
F(x)+F(2-x)=0, 
}
one notices that $F[R(N)]=-F[R(QQ)]$ and that the above $a$-function 
is the same as the one in the identical region 
in the original electric theory. Since the latter $a$-function 
are constructed via the prescription of \bound, one find that 
it is consistent with the electric-magnetic duality. 

As argued in Introduction, the auxiliary field method can be 
applied to the theories under consideration. In the original theory, 
turning on the superpotential 
$$
W=N_{ij}\left(Q^iQ^j-h\,M^{ij}\right), 
$$
to introduce the auxiliary fields $M^{ij}$ with the Lagrange multipliers 
$N_{ij}$, 
one can easily conceive that the parameter $h$ goes to zero in the infrared, 
due to the consistency with the result that the singlets $M^{ij}$ go to 
free fields in the magnetic theory for $7\leq{N_Q}\leq9$. 
Furthermore, when $h$ goes to zero, one obtains the superpotential $W_\ele$ 
of the other electric theory introduced above. It means that 
the original electric theory flows into the other electric theory with 
$W_\ele$ and thus is consistent with the magnetic picture. 
The equation of motion from the superpotential $W_\ele$ yields 
the constraint 
$$
{\partial\ \over \partial N_{ij}}W_\ele=Q^iQ^j=0.
$$
It is also consistent with the result that the composites $M^{ij}$ decouple from the remaining system in the original theory.

One may raise a question whether the auxiliary field method affects our results 
via $a$-maximization in the last section, because we introduced the auxiliary 
fields $M^{ij}$ and the Lagrange multipliers $N_{ij}$ charged under 
$U(1)\times{U(1)_R}$. This is however not the case, since 
as has been discussed in \athm, 
the massive fields $M^{ij}$ and $N_{ij}$ don't contribute to the $a$-function, 
due to \Fmassive. But, once the singlet $M^{ij}$ hits the unitarity bound, 
an accidental $U(1)_M$ symmetry appears to fix the $U(1)_R$ charge of $M^{ij}$ 
to $2/3$. On the other hand, the singlets $N_{ij}$ are still interacting with 
the vectors $Q^i$ in the superpotential, and their $U(1)_R$ charge remains 
unchanged and contributes $F[2-R(Q^iQ_j)]=-F[2R(Q)]$ to the $a$-function;
$$
F[R(M)]+F[R(N)] \quad\Rightarrow\quad F(2/3)+F[2-R(Q^iQ^j)]=-F[2R(Q)]+F_0.
$$
One can thus see that it gives the identical procedure to what we have done 
in the previous section when the meson $M^{ij}$ hits the unitarity bound. 

The auxiliary field method plausibly seems to work well to describe the flow 
of the original electric theory into the other electric one. We have seen that it gives a quite consistent picture with the magnetic one. 
However, if it is true, it seems to suggest a striking mechanism, 
because it means that the mass term $hN_{ij}M^{ij}$ goes to zero 
at the infrared fixed point. It is rather counter-intuitive and also is 
against {\it naturalness}. It would be interesting to inquire whether it could 
be a solution to the $\mu$ problem in the supersymmetric standard model\foot{
For a recent study of another phenomenological topic of this model, see 
\IINSY.}. 
Since the deeper implication of it is beyond the scope of this paper, 
we will have to leave it to the future. 

Although, in this paper, we restrict ourselves to the $Spin(10)$ gauge theory 
with a single spinor and $N_Q$ vectors, one can immediately extend it for 
the one with more than one spinor and $N_Q$ vectors \SpinX. 
%a wide class of supersymmetric gauge theories with magnetic dual description. 
We will report the results about it in another paper \KOTY, and we will also 
there give a detailed study of the deformations and the Higgs effect via 
$a$-maximization in the model discussed in this paper, as in \OkOo\ for 
the model discussed in \refs{\bound, \IWADE}.

%%%%%%%%%%%%%%%%%%%%%%%%%%%%%%%%%%%%%%%%%%%%%%%%%%%%%%%%%%%%%%%%%%%%%%%%%%%%%%
%\bigskip
\vskip 0.5in
\noindent
{\sl Note added}: after submission to the arXive, we notice that 
the gauge invariant operator $E_{1\alpha}$ is also redundant. 
Therefore, it doesn't invalidate our results in the region $VI$ for 
$N_Q=8,9$ and in the regions $VII$ and $VIII$ for $10\leq{N}_Q\leq21$. 
In order to prove that $E_{1\alpha}$ is redundant, 
as is discussed in \refs{\CDSW,\AddF}, we need to notice that, 
in the magnetic theory
$$
\bar{q}_j W_{\alpha}={1\over4}\bar{D}^2\left[D_{\alpha}
\left(\bar{q}_je^{-V}\right)e^V\right],
$$
where $W_{\alpha}=({1/4})\bar{D}^2[D_{\alpha}e^{-V}\cdot{}e^V]$ and 
$V$ is the vector superfield of the dual gauge group $SU(N_Q-5)$. 
Making use of the equation of motion 
$\partial{W_\mag}/\partial{s}=0$, we find that
$$\eqalign{
\left(*E_{1\alpha}\right)_{j_1\cdots{j}_{N_Q-7}}
&\sim \tilde{h}{\tilde\mu}^{N_Q-10}M^{kl}\bar{q}_k\bar{q}_{j_1}\cdots
\bar{q}_{j_{N_Q-7}}\left(\bar{q}_l\tilde{W}_{\alpha}\right)
\cr
&\sim {1\over4}\bar{D}^2\left[\tilde{h}{\tilde\mu}^{N_Q-10}M^{kl}
\bar{q}_k\bar{q}_{j_1}\cdots\bar{q}_{j_{N_Q-7}}
D_{\alpha}\left(\bar{q}_je^{-V}\right)e^V\right]. 
}$$
Thus, the Lorentz spinor $E_{1\alpha}$ is redundant.

%%%%%%%%%%%%%%%%%%%%%%%%%%%%%%%%%%%%%%%%%%%%%%%%
\break
%\bigskip\bigskip
%\vskip 1cm
%%%%%%%%%%%%%%%%%%%%%%%%%%%%%%%%%%%%%%%%%%%%%%%%
\centerline{\bf Acknowledgements}
We wish to acknowledge discussions with Masahiro Ibe, Takuya Okuda, and 
Yuuki Shinbara and to especially thank Yu Nakayama for helpful discussions 
on $a$-maximization and quantum moduli spaces. 
We are also grateful to Kenneth Intriligator for kind correspondence. T.~K. and Y.~O. would like to thank the participants of the Summer Institute String Theory 2005 at Sapporo, Japan for discussions. Y.~O. and F.~Y. thank the Yukawa Institute for Theoretical Physics at Kyoto University. Discussions during the YITP workshop YITP-W-05-08 on String Theory and Quantum Field Theory were useful to complete this work. The research of T.~K. was supported in part by the Grants-in-Aid (\#16740133) 
and (\#16081206) from the Ministry of Education, Science, Sports, and Culture 
of Japan. The research of Y.~O. was supported in part by JSPS Research 
Fellowships for Young Scientists. The research of Y.~T. was supported in part by the National Science Foundation under Grant No. PHY99-07949 and in part 
by JSPS Research Fellowships for Young Scientists. 

%%%%%%%%%%%%%%%%%%%%%%%%%%%%%%%%%%%%%%%%%%%%%%%%

\listrefs

\end